\documentclass[12pt]{iopart}

\usepackage{graphicx}%
\usepackage{multirow}%
\usepackage{amsmath,amssymb,amsfonts}%
\usepackage{placeins}
\usepackage{afterpage}

\usepackage{url}
\usepackage{amsthm}%
\usepackage{mathrsfs}%
\usepackage[title]{appendix}%
\usepackage{xcolor}%
\usepackage{textcomp}%
\usepackage{manyfoot}%
\usepackage{booktabs}%
\usepackage{algorithm}%
\usepackage{algorithmicx}%
\usepackage{algpseudocode}%
\usepackage{listings}%
\usepackage{textcmds}%
\usepackage{lipsum}
\usepackage{tabularx}
\usepackage{hyperref}
\usepackage{svg}
\usepackage[numbers]{natbib}
\usepackage{graphicx}
\usepackage{caption}
\usepackage{fancyhdr}
\def\bx{{\bf x}}
\def\bb{{\bf L}}
\def\bs{{\bf s}}

\def\bc{{\bf c}}
\begin{document}

\title[Efficient mapping of phase diagrams with conditional Boltzmann Generators]{\textbf{Efficient mapping of phase diagrams with conditional Boltzmann Generators}} 

\author{Maximilian Schebek$^1$, Michele Invernizzi$^2$,
Frank No\'e$^{1,2,3,4}$ and Jutta Rogal$^{1,5}$}

\address{$^1$ Fachbereich Physik, Freie Universit\"at Berlin,
14195 Berlin, Germany}
\address{$^2$ Fachbereich Mathematik und Informatik, Freie Universit\"at Berlin,
14195 Berlin, Germany}

\address{$^3$ Department of Chemistry,
Rice University, Houston, TX 77005, United States}

\address{$^4$ Microsoft Research AI4Science,
10178 Berlin, Germany}

\address{$^5$ Department of Chemistry, New York University,  New York, NY 10003, United States}
\ead{m@schebek@fu-berlin.de}

\vspace{10pt}
\begin{indented}
   \item[] \today
\end{indented}

\begin{abstract}
The accurate prediction of phase diagrams is of central importance for both the fundamental understanding of materials as well as for technological applications in material sciences. However, the computational prediction of the relative stability between phases based on their free energy is a daunting task, as traditional free energy estimators require a large amount of simulation data to obtain uncorrelated equilibrium samples over a grid of thermodynamic states. In this work, we develop deep generative machine learning models based on the Boltzmann Generator approach for entire phase diagrams, employing normalizing flows  conditioned on the thermodynamic states, e.g., temperature and pressure, that they map to. By training a single normalizing flow to transform the equilibrium distribution sampled at only one reference thermodynamic state to a wide range of target temperatures and pressures, we can efficiently generate equilibrium samples across the entire phase diagram. Using a permutation-equivariant architecture allows us, thereby, to treat solid and liquid phases on the same footing. We demonstrate our approach by predicting the  solid-liquid coexistence line for a Lennard-Jones system in excellent agreement with state-of-the-art free energy methods while significantly reducing the number of energy evaluations needed.
\end{abstract}

%
%
%
%
%

\section{Introduction}\label{sec1}
When studying or designing  materials, it is crucial to understand their phase diagrams, that is the relative stability of different phases as a function of temperature and pressure. Experimentally, the measurement of phase diagrams is challenging and often it remains uncertain whether the observed phases are truly the most stable ones and whether all possible phases have been discovered. 
The accurate and efficient prediction of phase diagrams is, therefore, one of the  central challenges in computational materials science~\cite{Chew2023-eo,Vega2008}. 

The determination of relative stability requires the evaluation of the free energies of all phases over the entire range of thermodynamic states of interest, which is a computationally complex and demanding task. 
Ideally, free energy calculations should be performed with \emph{ab initio} accuracy of interatomic interactions using large enough simulation cells and an exhaustive sampling of the phase space. Each of these factors, highly accurate interactions, large number of particles, and extensive sampling, can render the calculations prohibitively expensive and, in general, at least one of them needs to be compromised in order to make the approach computationally feasible. 
On the sampling side, an accurate estimate of the equilibrium distributions requires the exploration of the phase space with correct probability, which is generally achieved through the use of trajectory-based molecular simulation techniques such as Markov chain Monte Carlo (MC) and molecular dynamics (MD). These methods sample the equilibrium distribution by gradually changing the system configuration, thus requiring a large number of simulation steps for generating a sufficient number of statistically independent samples.   
Furthermore, traditional estimators for free energy differences,  such as free energy perturbation (FEP)~\cite{zwanzig_fep} and its multistate extension, the multistate Bennet acceptance ratio (MBAR)~\cite{Shirts2008-mbar}, require sufficient overlap in phase space between adjacent states for convergence. Therefore, a grid of simulation points needs to be defined over the relevant range of thermodynamic conditions, leading to a large number of simulations, all of which must be run until convergence. The required number of grid points for such thermodynamic integration is strongly system and state dependent, making the choice of a suitable grid a tedious trial-and-error procedure~\cite{Frenkel2001-yl}. 
 Targeted free energy perturbation (TFEP)~\cite{tfep} tackles the overlap problem by defining an invertible map on the configuration space. A suitable map increases the phase space overlap between states which reduces the number of required simulations points. However, defining a suitable map is far from trivial and TFEP has mostly been applied to simple problems where physical intuition was key to identifying such a map~\cite{Schieber2018ConfigurationalMS}.

In recent years, machine learning (ML) models have found broad use in molecular simulations. ML interatomic potentials have been shown to strongly reduce the cost of \emph{ab initio} MD simulations while retaining an accuracy comparable to density functional theory or other quantum chemistry methods~\cite{schnet_2018,Batzner2022,zhang_deep_potential_2018}. Lately, these potentials have also been applied  in the calculation of phase diagrams~\cite{niu_InitioPhase_2020,zhang_deep_water21} providing faster energy evaluations while the sampling challenge remains.
Complementary, deep generative models are being developed to generate uncorrelated samples of molecular structures in one go, with the aim of overcoming the sampling problem typical of MD and MC. 
In particular, the Boltzmann Generator (BG) approach~\cite{Noe2019} leverages a combination of highly expressive invertible coordinate transformations, such as normalizing flows~\cite{tabak_normflow,rezende_normflow}, and statistical mechanics methods such as importance sampling or Metropolis-Hastings Monte Carlo, to generate independent identically distributed samples of the targeted equilibrium distribution.
This approach has emerged both in the molecular sciences~\cite{Noe2019} as well as in lattice field theory~\cite{Albergo2019,Nikoli2021}. Within the TFEP-based  learned free energy perturbation (LFEP), using flows enables to learn flexible mappings for systems where physical intuition is insufficient~\cite{wirnsberger_lfep}. This approach has successfully been applied to sample small solvation systems~\cite{Wirnsberger_2022}, monoatomic solids~\cite{ahmad_free_2022,Wirnsberger_2023,van_leeuwen_boltzmann_2023}, and small molecules with complex quantum-mechanical potentials based on cheaper potentials~\cite{Rizzi2023} or at different temperatures to overcome slow modes~\cite{invernizzi_lrex}. However, while LFEP offers an elegant way to learn a mapping between two specific thermodynamic states, it cannot be used efficiently for the calculation of phase diagrams since a new flow model needs to be trained for each thermodynamic state.

In this work, we generalize the LFEP and BG approaches by employing normalizing flows conditioned on the target thermodynamic states. This conditional BG is capable of mapping MD samples from only one reference thermodynamic state across the entire $(T,P)$-range of the phase diagram with a single trained model. By training to match the equilibrium distributions at all thermodynamic states and appropriately reweighting the generated samples,
free energy differences can be very efficiently computed for any temperature and pressure. 
In this way, the coexistence line between phases can be mapped out with only minimal prior knowledge of its location, while avoiding the need to conduct numerous MD simulations across a grid of thermodynamic states. Using a permutation-equivariant architecture~\cite{wirnsberger_lfep}, our approach further treats ordered and disordered phases on the same footing. We demonstrate our developed framework by determining the coexistence line between solid and liquid phases of a Lennard-Jones (LJ) system consisting of 180 interacting particles, achieving highly accurate free energy estimates for both phases over the entire range of temperatures and pressures.

\section{Theoretical background}\label{sec:methods}

\subsection{Traditional free energy estimators}
We consider the isothermal-isobaric ($NPT$) ensemble, for which the reduced potential  of a  configuration $\bx$ is defined as~\cite{Shirts2008-mbar} 
\begin{align}\label{eq:u_red}
    u(\bx, V)&= \beta(U(\bx) + PV),
\end{align}
where $U(\bx)$ is the potential  energy, $V$ is the volume of the simulation box at configuration $\bx$, and $\beta= 1/k_BT$ denotes the inverse temperature. The reduced potential defines the equilibrium distribution
\begin{align}\label{eq:boltzmann}
    q(\bx,V) = Z^{-1}\exp(-u(\bx,V)),
\end{align}
where $Z=\int_\Gamma dVd\bx\,p(\bx,V)$ is the configurational partition function given by the integral over the configuration  space $\Gamma$.

From the partition function, the dimensionless configurational Gibbs free energy $f=\beta G$ can be obtained as
\begin{equation}\label{eq:f_logz}
    f =  - \log Z.
\end{equation}
Since evaluating the integral over configuration space is unfeasible for most interaction potentials, evaluating the absolute free energy is usually impossible. A more tractable approach consists in  evaluating the difference in free energy between two states $A$ and $B$, $\Delta f_{AB} = f_B - f_A $. In particular, many of the most widely used free energy estimators are based on the  Zwanzig identity~\cite{zwanzig_fep}, the central result of Free Energy Perturbation (FEP):
\begin{equation}\label{eq:f_diff}
     \Delta f_{AB} = - \log  \mathbb{E}_{(\bx,V)\sim q_A} \biggl[\exp(-\Delta u_{AB}(\bx,V))\biggr]
\end{equation}
Here, $\Delta u_{AB}(\bx,V) = u_B(\bx,V) - u_A(\bx,V)$ is the difference in the reduced potential between the two states. While being formally exact, the convergence of FEP crucially depends on the overlap between the corresponding configurational distributions $q_A(\bx,V)$ and $q_B(\bx,V)$, making it a highly biased and noisy estimator in practical calculations, even for small molecules~\cite{Mey_Allen_Bruce}.

A statistically more optimal estimator can be obtained using samples from both states of interest as done in the Bennett Acceptance Ratio (BAR) and its multistate extension MBAR~\cite{Shirts2008-mbar}. Nevertheless, the overlap problem remains which is commonly solved by defining a sequence of intermediate Hamiltonians between the two states of interest. However, such multi-staged approaches require samples from all intermediate states and it is not clear how best to define the intermediate stages.

Targeted free energy perturbation (TFEP)~\cite{tfep} tackles the problem of having intermediate states by defining an invertible map $\mathcal{M}$ transforming samples $(\bx,V)$ from $q_A$ to samples $(\bx',V')=\mathcal{M}(\bx,V)$ from a new target distribution $q'_{A}$ which shares a larger overlap with $q_B$ than $q_A$. For a perfect map, $q'_A=q_B$. Similarly, the inverse map $\mathcal{M}^{-1}$ transforms samples from $q_B$ to $q_B'$ and $q'_B=q_A$ for a perfect map. Using the Jacobian of the mapping, $J_\mathcal{M}(\bx,V)$, the generated distribution can be computed analytically using the change of variable theorem
\begin{equation}\label{eq:cov}
    q'_{A}(\bx',V') = q_A(\bx,V) |\det J_\mathcal{M}(\bx,V)|^{-1}.
\end{equation}
Samples drawn from the generated distribution can be reweighted to the target distribution by assigning a statistical weight $w$, which can be computed using various algorithms~\cite{Olehnovics2024}.
Here, we follow Ref.~\citenum{Noe2019} and define the importance weights of the generated samples as
\begin{equation}\label{eq:weight_def}
    w(\bx,V)\propto \frac{q_B(\mathcal{M}(\bx,V))}{q'_{A}(\mathcal{M}(\bx,V))}.
\end{equation}
The dimensionless free energy difference between $A$ and  $B$ can then be obtained as
\begin{equation}\label{eq:tfep}
    \Delta f_{AB} = -\log \mathbb{E}_{(\bx,V)\sim q_A}\bigl[w(\bx,V)\bigr].
\end{equation}
For the probability distributions of interest (Eq.\:\eqref{eq:boltzmann}), the importance weights are given by
\begin{equation}\label{eq:weight_npt}
    \begin{aligned}
        \log w(\bx,V) =\:&u_A(\bx,V)-u_B(\mathcal{M}(\bx,V)) \\
                          &  + \log|\det J_\mathcal{M}(\bx,V)|.
      \end{aligned}
\end{equation}
The weights can further be utilized to evaluate how well the learned distribution approximates the target distribution through the Kish effective sample size (ESS)~\cite{Kish1965-es}
\begin{equation}\label{eq:ess_kish}
    \text{ESS} = \frac{\bigl[\sum_i w(\bx_i,V_i) \bigr]^2}{\sum_i [w(\bx_i,V_i)]^2}.
\end{equation}
The ESS provides a rough  measure for how many uncorrelated samples would result in a Monte-Carlo estimator of the same statistical performance (typically reported relative to the total number of generated samples). 

The key idea of learned free energy perturbation (LFEP)~\cite{wirnsberger_lfep} consists in parametrizing $\mathcal{M}$ as a normalizing flow which removes the need of crafting a suitable mapping based on physical intuition only.

\subsection{Normalizing flows}\label{sec:flows}


Normalizing flows are a class of invertible deep generative models which are able to generate samples from a target distribution~\cite{tabak_normflow,papamakarios_flow}.  In contrast to other generative models, the use of invertible transformations allows to evaluate the generated distribution using the change-of-variables technique, making them an attractive sampling tool for the physical sciences. Normalizing flows aim to learn the optimal set of parameters $\theta^*$ of an invertible function $f(\bx)$ that maps samples $\bx$ from a prior distribution $q_A(\bx)$ into samples  from a target distribution $q_B(\bx)$. In practice, the learned set of parameters $\theta$ will be different from $\theta^*$ and the flow is trained by minimizing the Kullback-Leibler (KL) divergence between the generated distribution $q_A'(\bx)$ and $q_B(\bx)$. 

BGs target equilbirium distributions similar to Eq.~\eqref{eq:boltzmann}. In this case, the KL divergence can be expressed as~\cite{Noe2019,wirnsberger_lfep}:
\begin{equation}\label{eq:kl_div}
    D_{\rm KL}(q_A'||q_B) =  - \mathbb{E}_{\bx\sim q_A}\bigl[\log w(\bx)\bigr] - \Delta f_{AB},
\end{equation}
where the importance weights are defined similarly to Eq.\:\eqref{eq:weight_def} as $w(\bx)\!=\!q_B(f(\bx))/q'_{A}(f(\bx))$ and $\Delta f_{AB}$ is the free energy difference between $A$ and $B$. The loss function simplifies to 
\begin{equation}\label{eq:single_loss}
    \mathcal{L}_{AB}(\theta)  = - \mathbb{E}_{\bx\sim q_A}\bigl[\log w(\bx)\bigr] \geq \Delta f_{AB},
\end{equation}
since $\Delta f_{AB}$  is independent of the trainable weights.

For efficient training of normalizing flows, the determinant of the Jacobian of the transformation must be easy to calculate (see Eq.~\eqref{eq:weight_def}). For this purpose, coupling layers~\cite{dinh_nice_2014,dinh2017density} were designed which split the input data into two channels and transform one of the channels conditioned on the other. This yields a triangular Jacobian whose determinant can be evaluated analytically. The entire input vector can be transformed by stacking several coupling layers, whereby the channel to be transformed changes.

Flows can further be trained in a conditional setting which allows the modeling of conditional target distributions~\cite{ardizzone2020conditional,winkler_conditional_19}. A conditional normalizing flow  $f(\bx|\bc)$ is trained to learn a mapping between a prior distribution $q_A(\bx)$ and a conditional target distribution $q_B(\bx'|\bc)$ for a vector of conditioning variables $\bc$~\cite{ardizzone2020conditional,winkler_conditional_19}. In this case, the change of variable reads
\begin{equation}
    q'_A(\bx'|\bc) = q_A(\bx) |\det J_f(\bx|\bc)|^{-1}.
\end{equation}
For energy-based training, the trainable parts of the conditional KL divergence are given by the average of the KL divergences between the generated  and the target distributions conditioned on each target state over the entire range of $\bc$, yielding the following loss function:
\begin{align}\label{eq:cond_loss}
    \mathcal{L}(\theta) &  = - \mathbb{E}_{\bc\sim p_\bc}\mathbb{E}_{\bx\sim q_A}\bigl[\log w(\bx|\bc)\bigr]
\end{align}
with the conditional importance weights defined as  $w(\bx|\bc)\!=\!q_B(\bx|\bc) / q'_A(\bx|\bc)$. 
In practice, conditioning a coupling flow architecture can be easily achieved by concatenating the conditioning variable to the unchanged part before transforming the other part~\cite{ardizzone2020conditional}.  While originally developed in the context of image generation tasks~\cite{ardizzone2020conditional,winkler_conditional_19}, the idea of conditioning flows has recently been applied to the sampling of rare events and to lattice field theories~\cite{Falkner_2023, 10.21468/SciPostPhys.15.6.238}.

\subsection{Computing coexistence lines}\label{sec_method:coex_lines}
The coexistence line in the $(T,P)$ diagram of two phases of a system is defined by equal total Gibbs free energies $\tilde{G}(T,P)$. For a given state ($T_i, P_i$), $\tilde{G}_i$ is the sum of the configurational contribution (Eq.~\eqref{eq:f_logz})  and the kinetic contribution:
\begin{equation}
    \tilde{G}_i = G_i + \frac{3N}{\beta_{\rm i}} \log \Lambda(T_{\rm i}),
\end{equation}
where $\Lambda(T_i)$ is the thermal De-Broglie wavelength $\Lambda(T_i)=(\frac{h^2}{2\pi m k_B T_i})^{1/2}$. 

Using free energy estimators such as MBAR or TFEP,  the configurational contribution $G$ for each phase can be obtained by combining relative differences within phases with the absolute free energy $G_{\rm ref}$ at a reference state $(T_{\rm ref}, P_{\rm ref}$.)  Writing the difference in the dimensionless free energy between the reference point and a state $i$, $\Delta f_{{\rm ref},i}$ explicitly as
\begin{equation}\label{eq:abs_g}
    \Delta f_{{\rm ref},i} = \beta_iG_i - \beta_{\rm ref} G_{\rm ref},
\end{equation}
the absolute Gibbs free energy can be obtained as
\begin{equation}
    G_i = \frac{1}{\beta_i} \: \Delta f_{{\rm ref},i} + \frac{\beta_{\rm ref}}{\beta_i} G_{\rm ref}.
\end{equation}

$G_{\rm ref}$ is not directly available from simulations of the system of interest. Instead, it can be obtained by connecting the physical system to a model system whose Helmholtz free energy at the reference point, $F^0_{\rm ref}=F^0(T_{\rm ref},V_{\rm ref})$ with $V_{\rm ref}=\langle V\rangle_{T_{\rm ref},P_{\rm ref}}$, can be computed analytically. Then,  the reference free energy can be computed as
\begin{align}
    G_{\rm ref} = F^0_{\rm ref} + \Delta F^0_{\rm ref} + P_{\rm ref}V_{\rm ref}
\end{align}
with $\Delta F^0_{\rm ref}=F_{\rm ref} - F^0_{\rm ref}$ being the difference in Helmholtz free energy between the model and the physical system at the reference temperature. This quantity can be computed defining a set of $n_\lambda$ intermediate Hamiltonians between physical and model system for which $NVT$ MD simulations are performed. The resulting energy histograms serve as input for MBAR. In this work, we use as reference systems for the solid and the liquid phase the Einstein crystal~\cite{Frenkel2001-yl} and  the Uhlenbeck-Ford (UF) model~\cite{PaulaLeite2016}, respectively (see~SI for details).

\section{Conditional framework}
\subsection{Conditioning Boltzmann Generators
on temperature and pressure}\label{sec:tp_flow}

\begin{figure}[t]%
    \centering
    \vspace{.75cm}
    \includegraphics[width=\linewidth]{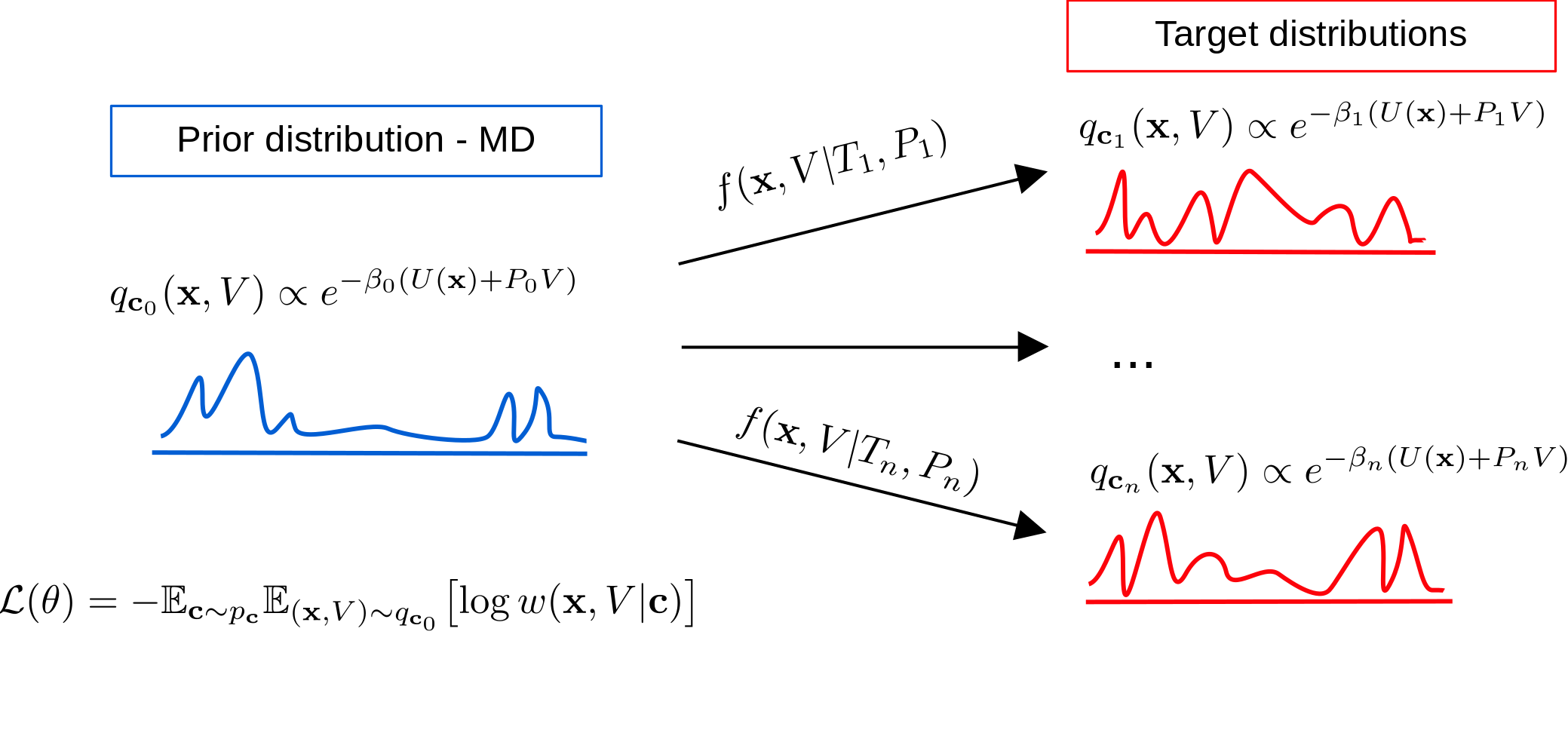}
    \vspace{-1.5cm}
    \caption{\textbf{Workflow conditional Boltzmann Generator.}  The  prior equilibrium distribution $q_{\bc_0}$ at a reference thermodynamic state $\bc_0=(T_0,P_0)$ is sampled with MD. A BG conditioned on the thermodynamic variables transforms this distribution to approximate the target equilibrium distributions $q_{\bc_i}$ at different thermodynamic states $\bc_i$. The BG is trained by minimizing the conditional loss function $\mathcal{L(\theta)}$, given as expectation value over the conditional value $\bc\sim p_\bc$ of the single-point KL divergences (Eq.~\eqref{eq:cond_loss_explicit}.}\label{fig:workflow}
\end{figure}

The key idea behind our approach is to compute free energy differences over a wide range of thermodynamic states without the need to run MD simulations at each state point.  This is achieved by training a BG conditioned on  temperature and pressure, which learns the mapping from a prior distribution at a reference thermodynamic state to a whole family of  target distributions across thermodynamic space.

Figure~\ref{fig:workflow} illustrates the workflow of our approach. We first sample the prior distribution at the reference state in the $NPT$ ensemble with MD and save samples consisting of the atomic configuration $\bx$ and the box volume $V$. In this work, we consider rectangular boxes and only allow for isotropic scaling, but the proposed approach is completely general and can easily be extended to fully triclinic simulation boxes.  Samples $(\bx,V)$ from the prior distribution  $q_{\bc_0}$ of the reference state $\bc_0=(T_0,P_0)$ with temperature $T_0$ and pressure $P_0$ 
are transformed by the flow to $(\bx',V')$ in order to approximate equilibrium distributions at multiple different thermodynamic conditions. 
Both transformations are conditioned on temperature and pressure, which is achieved by feeding the thermodynamic state as an additional input to the model during training (see Sec.~\ref{sec:architecture})~\cite{ardizzone2020conditional}.


To optimize the flow parameters, we employ conditional energy-based training (Eq.~\eqref{eq:cond_loss}), avoiding the need to sample the target distributions. Writing the conditional $NPT$ target equilibrium distribution as 
\begin{equation}
     q(\bx,V|\bc)\propto e^{-u(\bx,V|\bc)}
\end{equation}
with the reduced potential defined in Eq.~\eqref{eq:u_red} and adjusting the importance weights in Eq.~\eqref{eq:weight_npt} to the conditional setting, this yields the following loss function:
\begin{align}\label{eq:cond_loss_explicit}
    \mathcal{L}(\theta)  &  = - \mathbb{E}_{(T,P)\sim p_\bc}\mathbb{E}_{(\bx,V)\sim q_{\bc_0}}\bigl[\log w(\bx,V|\bc)\bigr]\\[5pt]
    & =  \mathbb{E}_{(T,P)\sim p_\bc}\mathbb{E}_{(\bx,V)\sim q_{\bc_0}}\bigl[  - \beta_0(U(\bx)  + P_0V)  +\beta(U(\bx')  + PV') \\[5pt]
            &    \hspace{3.9cm}               - \log|\det J_f(\bx,V|T,P) |\bigr] \nonumber.
\end{align}
The conditioning states $\bc\!=\!(T,P)$  are drawn from the distribution $p_\bc$, and  
$(\bx',V')$ is the sample generated by the conditional flow $f(\bx,V|T,P)$.
After training,  equilibrium expectation values of an observable $O(\bx,V)$ can be computed as
\begin{equation}\label{eq:reweight}
    \mathbb{E}_{(\bx,V)\sim q_\bc}\,\left[O(\bx,V)\right]\approx \frac{\mathbb{E}_{(\bx,V)\sim q_{\bc_0}}\left[w(\bx,V|\bc)O(\bx',V')\right]}{\mathbb{E}_{(\bx,V)\sim q_{\bc_0}}\left[w(\bx,V|\bc)\right]}.
\end{equation}
%
The free energy difference between the prior distribution and the target distribution is obtained from TFEP (Eq.\:\eqref{eq:tfep}) as
\begin{equation}\label{eq:tfep_flow}
    \Delta f_{\bc_0\bc} = -\log \mathbb{E}_{(\bx,V)\sim q_{\bc_0}}\bigl[w(\bx,V|\bc)\bigr].
\end{equation}
Compared to temperature steerable flows~\cite{Dibak2021}, the proposed workflow is more general and allows for more flexible models, as no requirements need to be placed on the architecture or the prior.

The evaluation of Eq.~\eqref{eq:cond_loss_explicit} involves the average of the KL divergences over multiple thermodynamic target states. One option to choose these target states is to define a set of discrete points which are used throughout training~\cite{ardizzone2020conditional,Falkner_2023}. However, for $n$ conditioning points this results in $n$ times the number of energy evaluations per sample. In addition, the spacing between the points has to be small enough such that the flow is able to efficiently interpolate between them.
We overcome these limitations by randomly drawing from a uniform distribution defined over the range of thermodynamic conditions of interest and transforming each sample with an individual $(T,P)$-value. This approach is computationally more efficient and also allows to condition over a continuous range rather than for a fixed number points.

\subsection{Flow architecture}\label{sec:architecture}
\begin{figure}[t]%
    \centering
    \includegraphics[width=\textwidth]{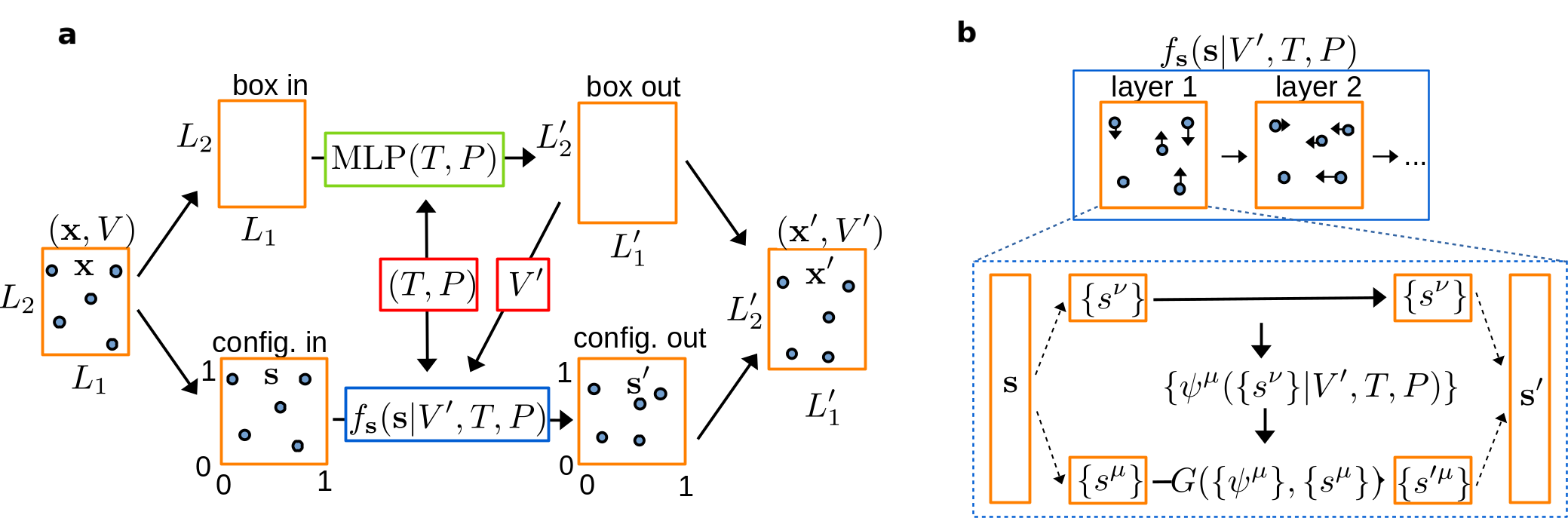}
    \caption{  \textbf{Flow transformation. a,} Schematic of the conditional transformation of one sample $f(\bx,V)=(\bx',V')$  consisting of an atomic configuration $\bx$ and a volume parameter $V$.  The transformation of the box parameter is a simple isotropic scaling operation ($L'_i/L'_j\!=\!L_i/L_j$) learned by an MLP conditioned on $(T,P)$. The configuration $\bx$ is scaled to fractional coordinates  $\bs$ and then transformed by a configurational coupling flow $f_\bs$ conditioned on $T$, $P$, and $V'$. As a final step, the configuration is scaled back to real coordinates. \textbf{b,} Illustration of the configurational flow transformation, where each flow layer transforms a subset of cartesian coordinates of all particles.  The dashed box is a sketch of one $(T,P,V')$- conditional coupling layer. The sets of fractional coordinates and spline parameters are denoted as $\{s\} = \{s_1,\dots, s_N\}$ and $\{\psi\} = \{\psi_1,\dots, \psi_N\}$, respectively (see Sec.~\ref{sec:layer} for details). Dashed lines denote splitting / merging operations.}\label{fig:architecture}
\end{figure}
The flow transforms a sample of the prior distribution $(\bx,V)$ to $(\bx',V')$, where the volume is connected to the box dimensions $\bb$ by  $V\!=\!\Pi_{i=1}^3 L_i$. In this work, we focus on the simulation of solids and liquids and therefore apply certain restrictions regarding the shape of the simulation box. We use orthorhombic simulation boxes with $\bb=[L_1,L_2,L_3]$ and further only allow for an isotropic scaling of the box ($L_i/L_j\!=\!{\rm const.}$), such that the box is fully characterized by its volume $V\!=\!\Pi_{i=1}^3 L_i$. 

Our flow is based on a shape-conditional architecture (see Fig.\:\ref{fig:architecture}~\textbf{a})~\cite{Wirnsberger_2023}. Here, the volume is updated by an affine transformation, while the atomic  configurations $\bx\in\mathbb{R}^{3N}$ are first scaled to fractional coordinates $\bs \in [0,1]^{3N}$  and then updated conditioned on the volume and the thermodynamic state. The volume transformation and the transformation of the position of particle $i$ can be written as
\begin{align}
    V' & = (1+\alpha(T,P))\cdot V + \beta(T,P),\label{eq:trafo_s} \\
    \bs_i & = \bx_i \oslash \bb, \label{eq:scale_init} \\ 
    \bs_i'& = f_{\bs,i}(\bs|V',T,P), \label{eq:trafo_coord} \\ 
    \bx_i'& = \bs_i' \odot \bb'.\label{eq:scale_final}
\end{align}
Here, $\alpha$ and $\beta$ are the output of a multilayer perceptron (MLP) taking as input pressure and temperature, where $\alpha$ is bounded below by -1. $\odot$ and $\oslash$ denote element-wise multiplication and division, respectively. The transformed box dimensions are given by  $\bb'=(V'/V)^{1/3}\bb$. 
$f_{\bs,i}$ is the output for particle $i$ of a $(V,T,P)$-conditional coupling flow over configurations (see Sec.~\ref{sec:layer}), operating on fractional coordinates. After the transformation, $\bs$ are transformed to physical coordinates by multiplying with the new box lengths.

The procedure just described gives rise to three contributions to the Jacobian, $J(V)$, $J_{\rm scale}(\bx)$, and $J(\bs)$. $J(V)$ arises from Eq.\:\eqref{eq:trafo_s} and is given by  $J(V) = \partial V'/ \partial V =\alpha(T,P) $. $J_{\rm scale}$ accounts for the initial global scaling of the  coordinates to fractional coordinates and the final scaling of the flow output to physical coordinates (Eqs.\:\eqref{eq:scale_init} and \eqref{eq:scale_final}) and is given by $J_{\rm scale}(\bx)=(V'/V)^N$. Finally, $J(\bs)=\partial \bs'/ \partial \bs $ accounts for the transformation in Eq.\:\eqref{eq:trafo_coord} and can be obtained from $f_\bs$.

\subsection{Conditional coupling layer}\label{sec:layer}
As condensed phase systems can exist in the liquid as well as in the solid phase, it is crucial to treat both phases on the same footing. Concretely,  $f_\bs$ must respect periodic boundary conditions and invariance under particle permutation. In addition, the center of mass needs to be fixed to prevent uncontrolled errors in the free energy prediction due to rigid translations~\cite{ahmad_free_2022}. 


To achieve these requirements, we extend the  coupling layer originally proposed in~\cite{wirnsberger_lfep} and refined in~\cite{Wirnsberger_2022} to the case of conditional sampling. Figure\:\ref{fig:architecture}~\textbf{b} shows our implementation of a  $(T,P,V)$ conditional coupling layer.  The $N$-particle configuration $\bs$  is split over cartesian coordinates into two subsets $\{s^\nu\}\!=\!\{s^\nu_1,...,s^\nu_N\}$ and $\{s^\mu\}\!=\!\{s^\mu_1,...,s^\mu_N\}$, e.g. $\nu\!=\!\!\{x\}$ and $\mu\!=\!\{y,z\}$. All coordinates indexed by $\nu$ remain unchanged, ${s'_i}^\nu = {s_i}^\nu$ , while the coordinates of atom $i$ carrying index $\mu$ are transformed according to
\begin{equation}
    {s'_i}^\mu = G(s_i^\mu,\psi_i^\mu),
\end{equation}
with
\begin{align}
    \psi_i^\mu = C^\mu_i(s^\nu_1,\dots, s^\nu_N|V',T,P).
\end{align}
Here, $G$ is a circular spline~\cite{rezende_flows_tori}, which naturally accounts for the periodicity of the problem. The parameters of the spline $\psi_i^\mu$ are obtained as the output of the network $C$ for the coordinates indexed by $\mu$ of particle $i$, which takes as input circular encodings of the coordinates $\nu$ concatenated with $T$,$P$, and $V'$. Following Refs.~\citenum{wirnsberger_lfep,Wirnsberger_2022}, $C$ is built as a specified number of transformer blocks each consisting of a multi-headed attention~\cite{vaswani2017attention} layer followed by an MLP. 
Both layers are implemented as residual updates. The partitioning of $\nu$ and $\mu$ is changed between layers such that each coordinate gets transformed (see~\cite{wirnsberger_lfep,Wirnsberger_2022} for details).

The invariance of the target density with respect to particle permutation is ensured due to  permutation-equivariance of the transformer update~\cite{kohler_equiv_flows}. In order to prevent the flow from learning a global shift, we fix one of the atoms and transform the coordinates of the remaining $N-1$ particles.

\section{Free energy differences and phase diagram}
\subsection{Training efficiency of conditioned and single-point normalizing flows}
\label{sec:training}
\begin{figure}[!ht]%
    \centering
    \vspace{-2cm}
    \includegraphics[width=\textwidth]{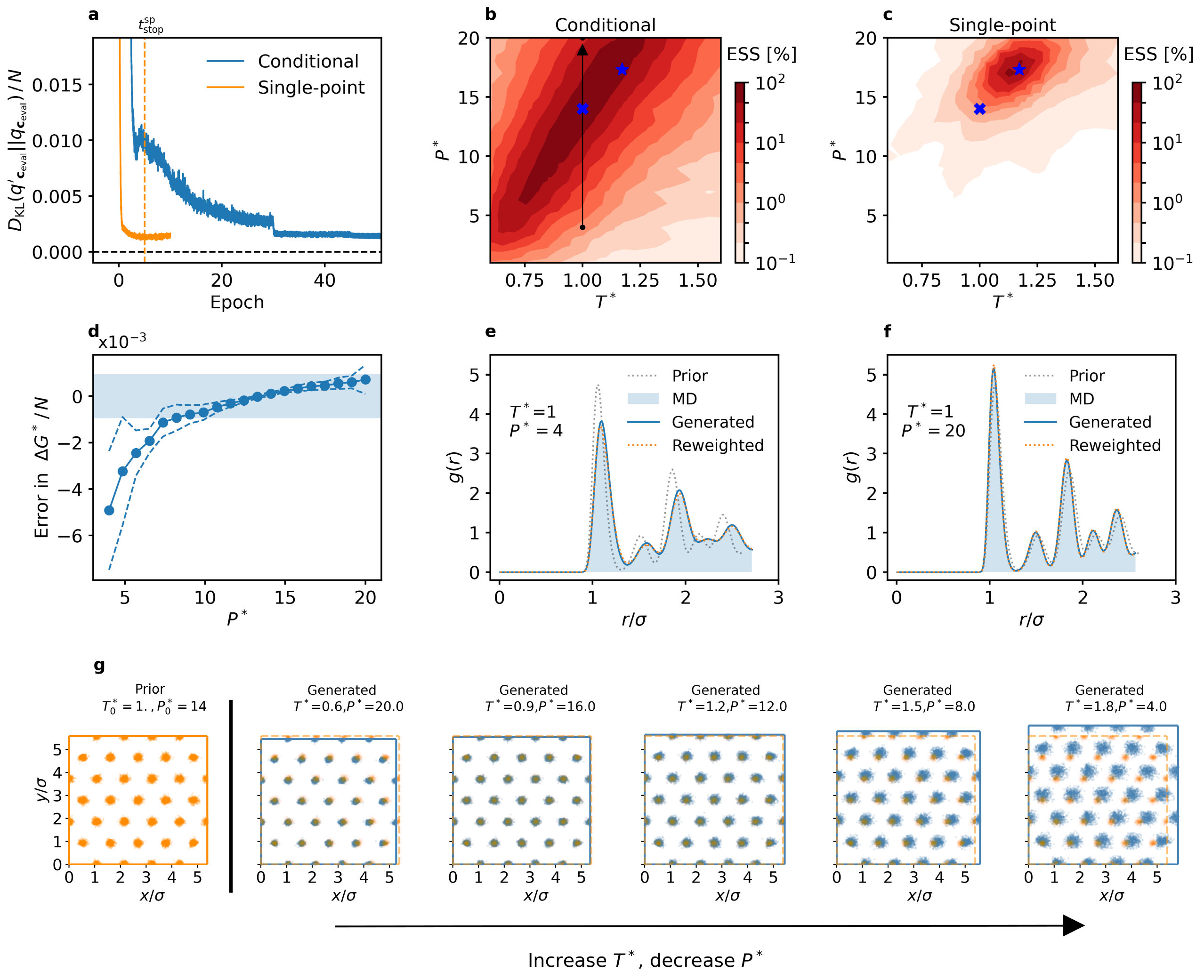}
    \vspace{-.5cm}
    \caption{\textbf{Training results for 180 Lennard-Jones particles in FCC phase. a, } KL divergence   (Eq.~\eqref{eq:kl_div}) between generated and target distributions at the evaluation state $\bc_{\rm eval}$ obtained from a flow trained only on the evaluation state (orange) and from a flow trained in a conditional way (blue). Thermodynamic states ($T^*_0,P^*_0$) and ($T^*_{\rm eval},P^*_{\rm eval}$) are marked as blue cross and blue star, respectively, in \textbf{b}. Learning rate reductions were applied for the conditional flow after 30 and 45 epochs. The orange dashed lines marks the training time after which the sampling efficiency of the single-point flow was evaluated.   \textbf{b} and \textbf{c}, 
     Sampling efficiency (Eq.~\eqref{eq:ess_kish}) for conditional and single-point flow, respectively, in \% evaluated over the range of thermodynamic states used for training the conditional flow. 
     \textbf{d,} Deviation in $\Delta G^*$ (relative to ($T^*_0,P^*_0$)) of the conditional flow from MD+MBAR along the path defined by the black arrow in \textbf{b}. The dashed lines indicate one standard deviation evaluated over 10 flow runs, the shaded area correspond to the maximum standard deviation of 10 MD+MBAR runs. 
     \textbf{e} and \textbf{f}, Radial distribution function of configurations from the prior, generated, and reweighted distributions of the conditional flow compared to MD at different thermodynamic states. \textbf{g}, 2D projections of 500 different configurations and mean box dimensions. The left-most panel contains samples from the prior distribution, the remaining panels show configurations generated for increasing temperature and decreasing pressure. For reference, prior configurations and mean box are superimposed to the generated configurations as orange shadows and dashed lines, respectively. In all plots, the fixed particle is located at (0,0).   \\
     }\label{fig:results_train}
\end{figure}
\afterpage{\FloatBarrier}

The conditioning framework introduced in the previous section has the advantage that samples can be generated over a continuous range of thermodynamic conditions. 
It requires, however, to train a conditional flow which might converge slower than a single-point flow trained only for one specific temperature and pressure. To compare the convergence of the two flow types, we train both a conditional and single-point flow  to transform samples of a 180 particle Lennard-Jones (LJ) face-centred cubic (FCC) crystal from a prior thermodynamic state at $\bc_0=(T^*_0,P^*_0)$ with $T^*_0\!=\!1.0, P^*_0\!=\!14.0$ (in reduced LJ units, see Supplementary Information (SI) for computational details). 
While the single-point flow is trained to generate samples of the equilibrium distribution at an evaluation point $\bc_{\rm eval}=(T^*_{\rm eval},P^*_{\rm eval})$ with $T^*_{\rm eval}=1.2, P^*_{\rm eval}=17$, the conditional flow is trained to generate samples for $T^* \in [0.6,1.6]$ and $P^* \in [1,20]$. Extensive statistics over the training process is collected by running 10 independent models (see SI for details).

 To measure the performances of the two flow types over training time, we evaluate the KL divergence (Eq.~\eqref{eq:kl_div}, with $\Delta f$ from standard MD+MBAR simulations) between the generated distribution and the target distribution at the evaluation point (Fig.~\ref{fig:results_train}~\textbf{a}). The KL divergence of the single-point BG  quickly converges within a few epochs, after which an increase in divergence indicates the onset of overfitting.   The KL divergence of the conditional BG is much more noisy  in the first 20 epochs of training,  with a temporary increase before continuing to decrease. The convergence behavior can be explained by the fact that the conditional BG requires more training steps as it learns a whole family of mappings simultaneously. After applying a learning rate reduction,  $D_{\text{KL}}$ becomes less noisy and converges to a similar value as the single-point BG. Even though training the conditional BG until convergence requires approximately 10 times more epochs than training the single-point BG, the additional training cost is amortized by the wide range of thermodynamic states covered by the conditional BG.

This becomes evident when comparing the statistical performance of the two flow types over the entire $(T,P)$-range, which is evaluated using the effective sampling size (ESS)~\cite{Kish1965-es} (Eq.~\eqref{eq:ess_kish}). The ESS over the entire $(T,P)$-range for the conditional and single-point flows  is shown in  Fig.~\ref{fig:results_train}~\textbf{b} and \textbf{c}, respectively.  While the single-point flow reaches high efficiencies close to the evaluation point, the efficiency quickly decays and is close to zero for all other $(T,P)$-conditions. In contrast, the conditional flow has a very high sampling efficiency of more  than 60\%  over a wide range of thermodynamic states. We found that within LFEP already a few tens of effective samples are sufficient to obtain an accurate estimate of  free energy differences. Within the  regions of largest  efficiencies, highly accurate free energy predictions can already be achieved evaluating only around  100 samples. Similarly, using 2\,000 test samples,  efficiencies as low as  1\,\% are sufficient to compute free energy differences with high accuracy.   
Remarkably, similar observations regarding sampling efficiency and free energy differences hold for the liquid phase, although the range of large sampling efficiencies was slightly smaller. For the corresponding efficiency map of the liquid, we refer to the~SI.

The sampling efficiency in Fig.~\ref{fig:results_train}~\textbf{b} is almost constant along a diagonal line in $(T,P)$ crossing the prior state. This may reflect the fact that along this line the flow has to learn only a small transformation, as  an increase in temperature leads to an increase in the atomic displacements and volume, whereas an increase in pressure has the opposite effect.

To assess the accuracy of the flow as free energy estimator, using 2\,000 test samples, we compute the Gibbs free energy difference $\Delta G^*$ (Eq.~\eqref{eq:tfep_flow}) with respect to the prior along the path  $P^*\in[4,20]$ for a fixed $T^*=1.0$ and evaluate the error compared to results obtained with MBAR used across a set of MD  simulations (Fig.~\ref{fig:results_train}~\textbf{d}). Overall,  we observe that both methods are in excellent agreement over the entire range.
The accuracy only decreases slightly with increasing distance from the prior and decreasing sampling efficiency. Nevertheless, even for the furthest point, $P^*=4$, the mean deviation per particle is only $5\cdot10^{-3}$, while the absolute value in $\Delta G^*$ between $P^*=4$ and $P^*=20$  is around 14.5  per particle, 
that is the maximum relative error along the line is less than 0.05\%.
This demonstrates that the conditional BG can be used for the efficient evaluation of free energy differences also for states far away from the prior.

Furthermore, we illustrate the ability of the model to  accurately capture pairwise correlations over varying thermodynamic conditions by computing the radial distribution function (RDF) $g(r)$~\cite{Frenkel2001-yl} for configurations at the start and end point of the path  (Fig.~\ref{fig:results_train}~\textbf{e} and \textbf{f}) and compare to MD. 
For both $(T,P)$-conditions, the RDFs resulting from samples of the generated distributions  are nearly indistinguishable from the MD results, even before reweighting, while samples from the prior distribution clearly yield  different RDFs.  Any remaining deviations between the mapped and MD results are 
resolved by reweighting the generated samples (Eq.~\eqref{eq:reweight}).  The ability of the conditional BG model to generate samples with varying displacements and volumes is further illustrated in Fig.~\ref{fig:results_train}~\textbf{g}, where 2D-projections of 500 generated configurations with increasing temperature and decreasing pressure are shown.

\subsection{Mapping out coexistence lines}\label{sec:coex_line}

\begin{figure*}[t]%
    \centering
    \includegraphics[width=\textwidth]{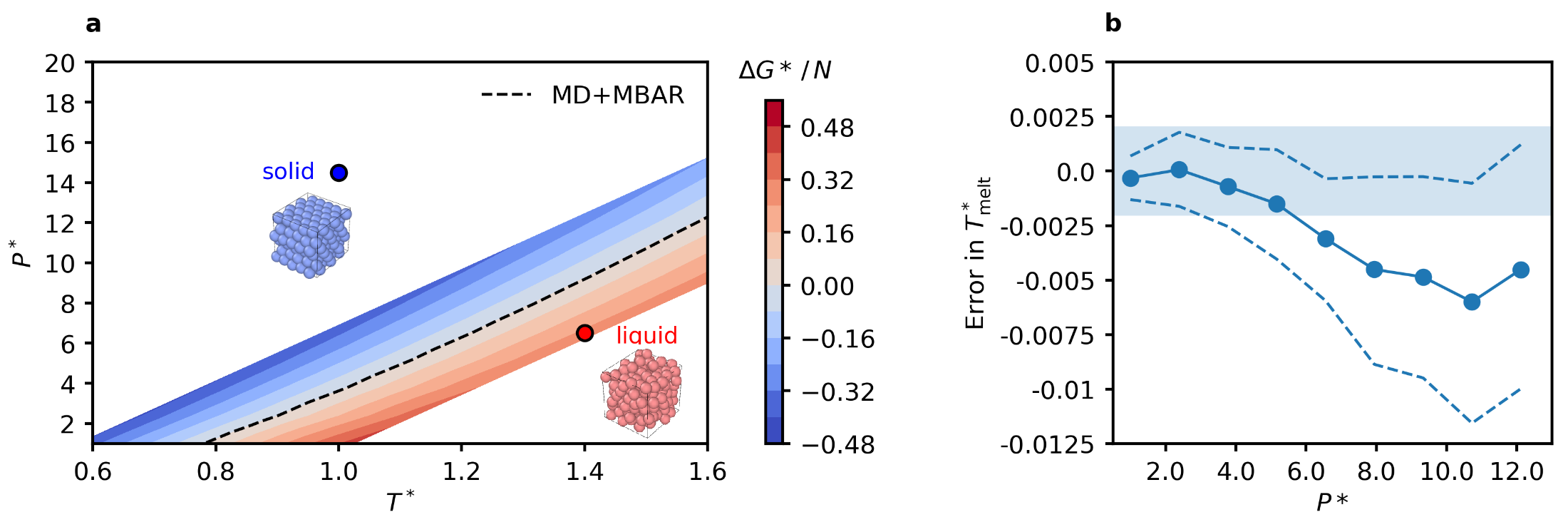}
    \caption{\textbf{Coexistence line a,} Contourplot of $\Delta G^* = G^*_{\rm sol} - G^*_{\rm liq} $  per particle as obtained from the conditional BGs. Liquid-solid coexistence corresponds to $\Delta G^*\!=\!0$. The dashed line denotes the coexistence line  obtained from a high-accuracy MD+MBAR run. The locations of solid and liquid priors are marked as blue and red circles, respectively.  \textbf{b,}  Deviation in the melting temperature of the conditional BG from MD+MBAR.  The uncertainties of the BG predictions were obtained as the standard deviation over 10 independent runs. The shaded area indicates the maximum fluctuations over 10 MD+MBAR runs.}\label{fig:phase_diag}
\end{figure*}

Having demonstrated the efficiency of the conditional BG for accurately computing free energy differences over a wide range of thermodynamic states, we now apply this methodology to the exploration of the phase diagram of the  LJ  system.

Determining the coexistence line requires to find the thermodynamic states of equal free energy. As discussed in the introduction,
this is  a tedious and expensive task using traditional free energy estimators as a suitable set of discrete thermodynamic states needs to be identified and  full MD/MC simulations, including potentially large equilibration times, need to be executed at each point~\cite{Chodera2016-lq}. In the case of liquid-solid coexistence lines, an additional complication arises from the fact that suitable states, in which both the solid and the liquid phase can be simulated and no phase transition occurs during the simulation time, need to be selected by hand.

Using the LJ system as an example, we demonstrate that our framework based on $(T,P)$-conditioned flows can alleviate these issues.  Specifically, we compute the coexistence line between the solid FCC  and liquid phase. The only user-defined input to our workflow 
consists in choosing appropriate locations of the prior states for the two phases of interest. If a point of coexistence is already known, this point is a good choice as the sampling efficiency is  highest close to the prior location (see Fig.~\ref{fig:results_train}) which allows to compute the coexistence line with only a small number of samples. 
To emphasize that our approach works without explicit knowledge of the coexistence region, we define a broad range of thermodynamic states with $T^* \in [0.6,1.6]$ and $P^* \in [1,20]$ in which the coexistence line is expected~\cite{bizjak_lj_2008}. Based on physical intuition, we place the prior for the solid at low temperature and high pressure $(T^*_{\rm sol}=1, P^*_{\rm sol}=14)$, and for the liquid at high temperature and low pressure $(T^*_{\rm liq}=1.4, P^*_{\rm liq}=6)$. Conditional flows for both phases are trained over the expected region of coexistence (details are given in the~SI). After training, we define a grid of points in the $(T^*,P^*)$ diagram with a spacing of $(\Delta T^*,\Delta P^*)$=(0.05,1.4) and use the BGs to 
evaluate the relative free energies to the respective prior based on 2000 samples.

To obtain the difference in free energy between the two phases, the absolute free energies at the prior locations are required. 
These are determined by evaluating the free energy difference to analytically or numerically tractable reference systems. 
For the solid phase, this could also be achieved using normalizing flows, as shown in Ref.~\citenum{Wirnsberger_2023}.
Here, we instead chose the more traditional and computationally more efficient approach of 
MBAR together with MD simulations.
For the solid phase, the Einstein crystal~\cite{Vega2008} serves as a reference, while the liquid phase is connected to the Uhlenbeck-Ford model~\cite{PaulaLeite2016}  (see Sec.~\ref{sec_method:coex_lines} for details).

In Fig.~\ref{fig:phase_diag} \textbf{a}, the free energy difference between the solid and liquid phase around the coexistence line is shown as a function of temperature and pressure (see~SI for free energy differences over the entire $(T,P)$-range considered during training).  The coexistence line predicted by the BG is in excellent agreement with the reference values, obtained from high-accuracy MD+MBAR simulations (dashed line in Fig.~\ref{fig:phase_diag} \textbf{a}), with only small deviations in the high-temperature region.
In Fig.~\ref{fig:phase_diag} \textbf{b}, the error in the melting temperatures $T^*_{\rm melt}$ as a function of pressure are shown as determined by the BG in comparison to MD+MBAR. For the low-pressure regions, the melting temperatures predicted by the BG are well within the error bars of the reference calculations. Only for higher pressures, deviations become slightly larger. Overall,   the point of equal free energy, and therefore the melting temperatures, can be determined with a maximum (mean) deviation accuracy of less than $0.01\,T^*$.

\subsection{Computational efficiency} 
A direct comparison of the computational costs for mapping out an entire phase diagram is not entirely straightforward as it depends on a variety of factors.
Here, we focus on the number of required energy evaluations which will dominate 
our flow-based approach as well as traditional free energy estimators for more accurate and expensive interaction models, including ML potentials and \emph{ab initio} approaches.

For MBAR, the number of energy evaluations is determined by the discretization of the $(T,P)$-range and the number of samples needed for convergence at a given grid density. For the BG, generating a sufficient number of prior samples constitutes the main computational investment.  The LJ solid in this study melts when the coexistence line is overstepped by around 0.1\,$T^*$ or 1\,$P^*$. 
Consequently, the grid spacing for MBAR should  be at maximum  $(\Delta T^*,\Delta P^*)$=(0.05\,$T^*$,0.5$\,P^*$) resulting in a total of 800 simulation points per phase. At this grid spacing, at least 100 MD samples per state separated by 500 decorrelation steps  need to be collected for MBAR to converge the coexistence line with similar accuracy as the BG, 
where each MD simulation additionally requires at least 20\,000 initial equilibration steps. In total, this amounts to 56 million
energy evaluations per phase.  For the chosen flow architecture, including more than 20\,000 prior samples does not substantially improve the sampling efficiency. Generating these MD samples with the same number of decorrelation steps  requires around 10 million energy evaluations, while training for 50 epochs adds less than one million. The final evaluation of 2000 samples on the grid of 800 points adds 1.6 million
energy evaluations. Within this rough estimate, the number of required energy evaluation is already reduced by a factor of 5 for the system presented here.
In addition, flow-based generation of samples  does not require the evaluation of forces which, for expensive potentials, constitutes an additional advantage.


\section{Discussion}
The introduced Boltzmann Generator framework based on conditional normalizing flows provides a novel approach to efficiently calculate differences in free energies over a wide range of thermodynamic states. Contrary to traditional approaches, such as thermodynamic integration or expanded ensembles, it does not require an explicit discretization of the temperature and pressure space.
For both solid and liquid phases, flows trained in a conditional fashion achieve large sampling efficiencies even far away from the prior states.
As a consequence, only minimal prior knowledge about the system and the melting region is required to map out coexistence lines between two  phases with high accuracy. Our workflow, therefore, constitutes a first step towards the efficient and simple calculation of phase diagrams using normalizing flows.

 The combination of MD prior data with normalizing flows and the subsequent connection to tractable reference systems allows to compute $(T,P)$-dependent absolute free energies of the solid phase much more efficiently than purely flow-based approaches~\cite{Wirnsberger_2022,Wirnsberger_2023}. In addition, our approach is equally applicable to evaluate absolute free energies of the liquid phase which is currently out of reach with purely flow-based methods~\cite{coretti_learning_2022}. Nevertheless, the connection to the tractable reference systems remains computationally expensive. 
 As a next step, a flow-based framework to directly compute free energy differences between phases, ideally utilizing the prior samples of the two phases, would be desirable. 
 
 A second direction concerns the system sizes which can be tackled with our approach. In our studies as well as in previous research~\cite{abbott_aspects_2022}, it was found that the sampling efficiency decreases significantly with increasing system sizes. For the application to realistic systems, thousands of atoms may be required to accurately reproduce experimental results which is unfeasible using current flow-based sampling methodologies. A possible solution could consist in training a transferable model on small systems that can then efficiently be applied to larger ones. With an increase in sampling efficiency, the application of our approach to more accurate potentials as well as the inclusion of molecular crystals~\cite{kohler2023rigid} becomes feasible and is planned in future work. By advancing the aforementioned fields, we aim to improve the utility and applicability of our approach to enhance the understanding of complex phase behaviour in realistic systems.


\section*{Code availability}
The flow models developed in this work are available on GitHub at
\url{https://github.com/maxschebek/flow_diagrams}. Models were built and trained using JAX~\cite{jax2018github}, equinox~\cite{kidger2021equinox}, and jax-md~\cite{jaxmd2020}.

\section*{Acknowledgements}
MS and JR acknowledge financial support from Deutsche Forschungsgemeinschaft (DFG) through grant CRC 1114 \ldq Scaling Cascades in Complex Systems\rdq, Project Number 235221301, Project B08 \ldq Multiscale Boltzmann Generators\rdq.  JR acknowledges financial support from DFG through the Heisenberg Programme project 428315600. MI acknowledges support from the Humboldt Foundation for a Postdoctoral Research Fellowship. We thank Leon Klein for insightful discussions.

 \setcounter{table}{0}
\setcounter{figure}{0}
\renewcommand{\thefigure}{S\arabic{figure}}
\renewcommand{\theequation}{S\arabic{equation}}
\renewcommand{\thetable}{S\arabic{table}}
\setcounter{equation}{0}
\setcounter{table}{0}
\setcounter{section}{0}
\renewcommand\thesection{\Alph{section}}

\newcommand{\newblock}{}

\bibliographystyle{unsrt}
\bibliography{article.bib}
 \setcounter{table}{0}
        \setcounter{figure}{0}
        \renewcommand{\thefigure}{S\arabic{figure}}
        \renewcommand{\theequation}{S\arabic{equation}}
        \renewcommand{\thetable}{S\arabic{table}}
        \setcounter{equation}{0}
        \setcounter{table}{0}
        \setcounter{section}{0}
        \renewcommand\thesection{\Alph{section}}

\setcounter{section}{0}

\newpage

{\noindent\Large{\textbf{Supplementary Information}}}
\section{Training details}
\begin{table}[h]
    \setlength\tabcolsep{20pt}
    \caption{Hyperparameters used for training. \label{tab:params_flow}}
    \begin{tabularx}{\textwidth}{l l}
        \hline        \hline\\[-1.5ex]
    \textbf{Configurational flow}\\
    Number of layers & 12 \\ 
    Number of transformer blocks & 4 \\ 
    Number of transformer heads & 4 \\ 
    Number of frequencies for circular encoding & 8\\ 
    Number of spline segments & 16\\
    Embedding dimension & 128\\
    Number of nodes in MLP & 128\\
    Number of layers in MLP & 2\\
    \\
    \textbf{Shape update}\\
    Number of nodes in MLP & 16\\
    Number of layers in MLP & 2\\
    \\
        \textbf{Optimization}\\
    Initial learning rate  &  $10^{-4}$\\
    Number of epochs  &  50\\
    Learning rate decay  &  .1\\
    Learning rate decay steps  &  30 epochs, 45 epochs\\
    Optimizer  &  Adam ~\cite{kingma_adam}\\
    Batch size  &  128\\
     \hline\hline
     \end{tabularx}

\end{table}

Table\:\ref{tab:params_flow} lists the parameters used for training the flows. We reduced the learning rate for the training of the conditional flows after 30 and 45 epochs by factor .1 to guarantee a converging loss. All presented flow results (sampling efficiencies and free energy differences) are the average over 10 different runs with differently initialized models, where each run employed an independent set of prior samples. The prior samples were generated by running 10 MD simulations and collecting a total of 20\,000 samples per simulation which were then randomly shuffled and divided into 10 sets of 20\,000 samples. Configurations and box dimensions were stored every 500 steps and  each simulation was started by 500\,000 steps of equilibration time. The prior samples were further randomly divided into training sets consisting of and test sets consisting of 18\,000 and 2\,000 samples, respectively. The unseen test sets were used for the calculation of free energy differences to ensure convergence to the correct value~\cite{Rizzi2021}.
The flow models trained around 40 minutes on a single NVIDIA RTX A5000 GPU and drawing 10\,000 samples samples from the model took around 4 seconds.

\section{Simulation details}

All MD simulations have been carried out using the OpenMM package~\cite{Eastman2017} using the standard 12-6 Lennard-Jones pair potential given by
\begin{align}
    U_{\rm LJ}(r) = 4\varepsilon\Biggl[\biggl(\frac{\sigma}{r}\biggr)^{\!\!12} - \biggl(\frac{\sigma}{r}\biggr)^{\!\!6} \Biggr],
\end{align}
where the distance between two particles is denoted as $r=|\bx_1-\bx_2|$ and $\varepsilon$ and $\sigma$ define the units of energy and length, respectively. In these units (also called reduced units), reduced temperature and pressure, $T^*$  and $P^*$, can be connected to real units via $T^*=k_BT/\varepsilon$ and $P^*=P\sigma^3/\varepsilon$ (see Tab.\:\ref{tab:params} for values used in the MD simulations).

For the FCC structure, 180 particles\ were set up with 6 close-packed ($x-y$) layers along the [111] direction, each consisting of 30 atoms. Simulations of the liquid were conducted by first melting the FCC structure using a high temperature and then reducing the temperature to the target value. We used a switching function to obtain a smoothly decaying potential between the switching radius $r_{\rm switch}$ and the cutoff radius $r_c$. For $r_{\rm switch}<r<r_c$, the energy is multiplied by~\cite{Eastman2017}
\begin{equation}
    S = 1 - 6x^5 + 15x^4 - 10x^3,
\end{equation}
where $x=(r - r_{\rm switch})/(r_c - r_{\rm switch})$.

For all simulations, we used a Langevin integrator, for the $NPT$ simulations an isotropic Monte-Carlo barostat was used as implemented in OpenMM. The reference potentials for the Einstein crystal and the Uhlenbeck-Ford model were implemented as custom potentials in OpenMM.

All MBAR results were obtained using 1\,000 MD samples per state which were collected every 500 steps to ensure decorrelation. Each simulation was started by 500\,000 steps of equilibration time. Similar to the flow results, uncertainties of the MD+MBAR predictions were obtained over 10 independent simulations.  The grid of thermodynamic states was chosen to be the same as for the respective flow results. 

\section{Reference potentials}

The Einstein crystal is defined as a system of $N$ non-interacting harmonic oscillators with energy
\begin{align}
    U_{\rm EC}(\bx_1,\dots,\bx_N) = \Lambda_{\rm EC} \sum_{i=1}^{N} |\bx_i - \bx_i^0|^2,
\end{align}
where the equilibrium position of atom $i$ is denoted as $\bx_i^0$ and $\Lambda_{\rm EC}$ is the spring constant. The reduced absolute free energy including both the configurational and kinetic contribution of the EC crystal $\tilde{f}_0=\beta \tilde{F}_0$ can be computed as~\cite{Vega2008,polson2000}
\begin{align}
    \tilde{f}_0 / N  = \frac{1}{N}\ln\biggl(\frac{N\Lambda^3}{V}\biggr) +  \frac{3}{2}\biggl(1-\frac{1}{N}\biggr)\ln\biggl(\frac{\beta\Lambda_{\rm EC}\Lambda^2}{\pi}\biggr) - \frac{3}{2N}\ln N. \nonumber
\end{align}
Here, $\Lambda$ denotes the thermal De-Broglie wavelength. Following~\cite{Vega2008}, $\Lambda$ is set to the Lennard-Jones parameter $\sigma$, leading to a temperature-independent momentum contribution to the free energy, which, consequently, vanishes in the computation of free energy differences.

\begin{table}[t]
    \caption{Parameters used for the different potentials. \label{tab:params}}
    \setlength{\tabcolsep}{20pt}
    \renewcommand{\arraystretch}{1}
    \begin{tabularx}{\textwidth}{lXXXXXXX}
     \hline        \hline\\[-1ex]
    $\varepsilon$&$\sigma$ & $r_c$&  $r_{\rm switch}$&  $\Lambda_{\rm EC}$ & $\Lambda$ & $p$ & $\sigma_{\rm UF}$ \\ [.1cm]
     \hline\\[-1.5ex]
     0.99\,$\frac{\rm kJ}{\rm mol}$&0.34\,nm& 2.2\,$\sigma$& 0.9\,$r_c$ & 2500\,$\frac{\rm kJ / mol}{\sigma^2}$ & $\sigma$ & 50 & $0.58\,\sigma$ \\[.1cm] 
     \hline\hline
     \end{tabularx}

\end{table}

The (scaled) Uhlenbeck-Ford model~\cite{PaulaLeite2016} serves as a reference potential for the liquid phase. It is given as a pair potential defined by
\begin{equation}
    U^{(p)}_{\rm UF} (r) = -\frac{p}{\beta}\ln\biggl(1-e^{-(r / \sigma_{\rm UF})^2}\biggr),
\end{equation}
where $p$ is an integer and $\sigma_{\rm UF}$ is a length parameter. In our calculations, we used $p=50$ and $\sigma_{\rm UF}=0.58 \sigma$. We used the same cutoff as for the LJ potential. The reduced excess free energy $f^{\rm exc}_{\rm UF}=f_{\rm UF} - f_{\rm IG}$, where $f_{\rm IG}$ denotes the free energy of the ideal gas, of the UF model can be computed as~\cite{PaulaLeite2016}
\begin{align}\label{eq:uf_deltaf}
    \frac{f^{\rm exc}_{\rm UF}(x) }{N} =  \sum_{n=1}^\infty\frac{\tilde{B}^{(p)}_n}{n}x^n,
\end{align}
with  $x\!=\!b\rho$ and $b\!=\!1/2(\pi\sigma_{\rm UF}^2)^{(3/2)}$. $\tilde{B}^{(p)}_n\!=\!B^{(p)}_n / b^{n-1}$ can be computed in terms of the virial coefficients $B^{(p)}_n$, $\rho$ is the particle density. To compute Eq.\:\eqref{eq:uf_deltaf}, we make use of the \verb+ufGenerator.py+ script provided in~\cite{PaulaLeite2016}.

To compute the absolute free energy of the liquid phase, we further need the reduced free energy including the kinetic contribution of the ideal gas (IG) which is given by~\cite{Vega2008}
\begin{equation}\label{eq:ideal_gas}
    \tilde{f}_{\rm IG} / N = \ln\biggl(\frac{N\Lambda^3}{V}\biggr)- 1 + \frac{\ln(2\pi N)}{2N}.
\end{equation}
Similar to the solid, we set $\Lambda=\sigma$, thus neglecting the temperature dependence of the momentum contribution to the free energy.

In order to compute the absolute free energy of the Lennard-Jones system, we connect it to the respective reference system using a set of intermediate potentials via
\begin{align}
    U(\lambda) = \lambda U_{\rm LJ} + (1-\lambda)U_{\rm ref},
\end{align}
where $\lambda\in[0,1]$ is an interpolation parameter and $U_{\rm ref}$ is the EC potential if the system is in a solid phase or the UF model if the system is liquid. We used $n_\lambda=100$ equidistant intermediate potentials for which MD simulations are performed. For each simulation, we collect $10^3$ samples which are used as input for MBAR~\cite{Shirts2008-mbar}.


\section{Efficiency map fluid phase}
\begin{figure*}[!h]%
    \centering
    \includegraphics[width=.5\textwidth]{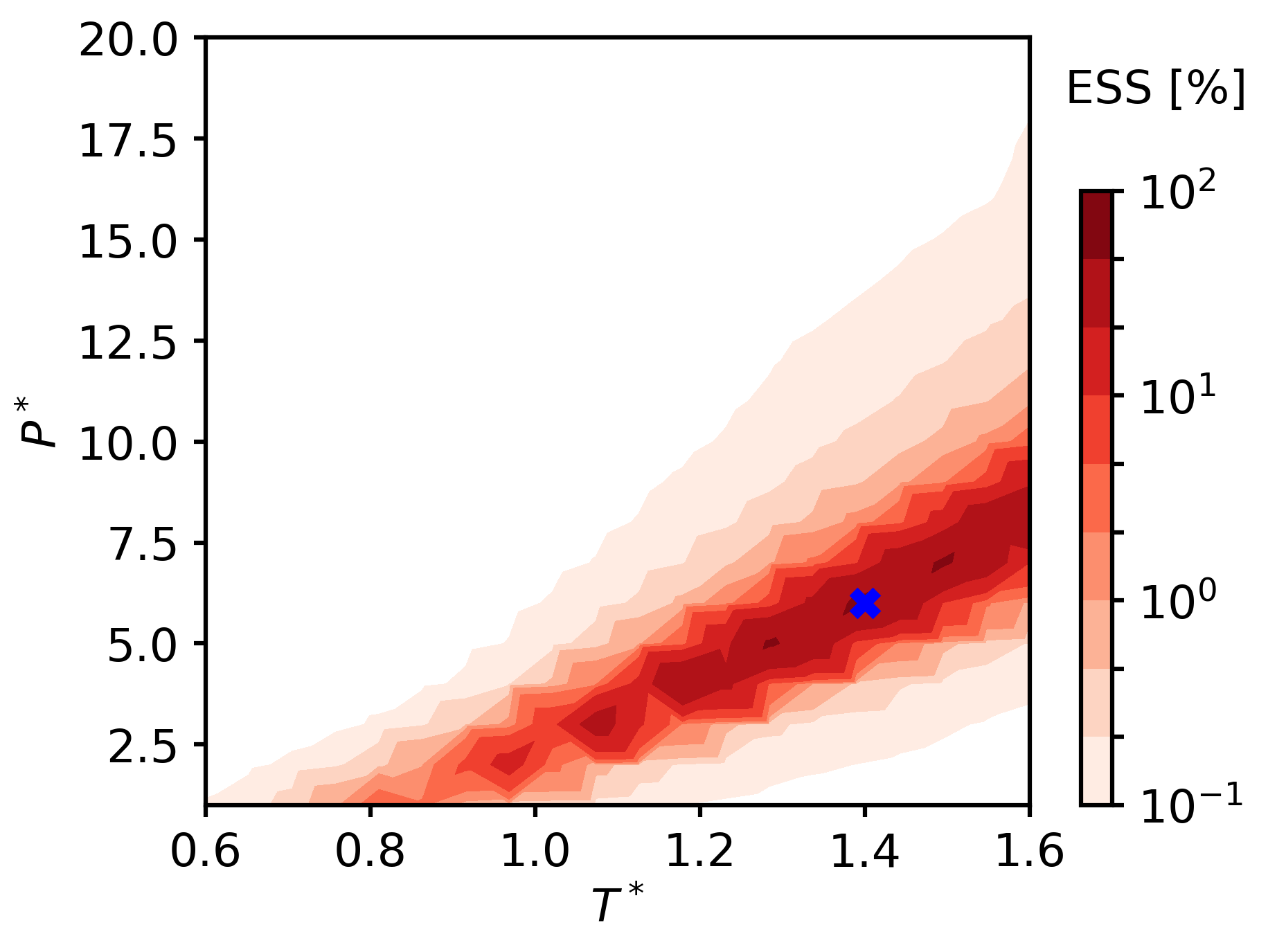}
    \caption{Sampling efficiency of the conditional flow for the fluid phase in \%. The blue cross denotes the prior location $T^* = 1.4,P^* = 6$.  }\label{fig:effmap_fluid}
\end{figure*}

Figure~\ref{fig:effmap_fluid} shows the sampling efficiency of the conditional flow for the fluid phase of the LJ potential. Training parameters were same as for the solid phase.

\clearpage
\section{Free energy differences}
\begin{figure*}[!h]%
    \centering
    \includegraphics[width=.5\textwidth]{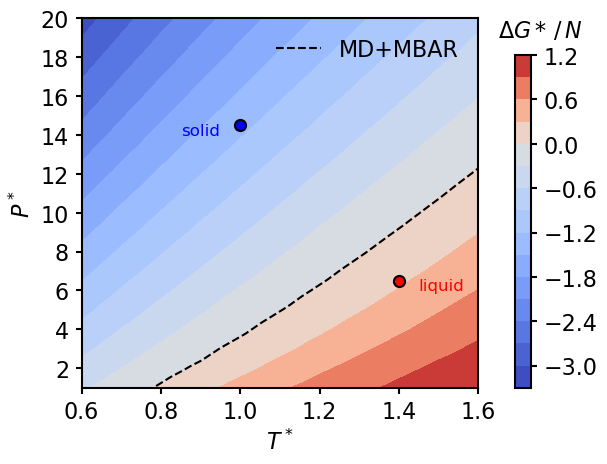}
       \caption{ Contourplot of $\Delta G^* = G^*_{\rm sol} - G^*_{\rm liq} $  per particle as obtained from the conditional flows. Liquid-solid coexistence corresponds to $\Delta G^*\!=\!0$. The dashed line denotes the coexistence line  obtained from a high-accuracy MD+MBAR run. The locations of solid and liquid priors are marked as blue and red circles, respectively.}\label{fig:phase_diag_full}
\end{figure*}

Figure~\ref{fig:phase_diag_full} shows the free energy differences obtained using the conditional flows over the whole range of thermodynamic states considered for training.

\end{document}